\documentclass[twocolumn,aps,pra,notitlepage,amsmath,amssymb,superscriptaddress]{revtex4-1} 

\usepackage{graphicx}
\usepackage{dcolumn}
\usepackage{bm}
\usepackage{bm}
\usepackage{amsmath}
\usepackage{verbatim}     
\usepackage[dvipsnames]{xcolor}       
\usepackage[normalem]{ulem}   
\usepackage{bbold}
\usepackage[utf8]{inputenc}
\usepackage[T1]{fontenc}

\usepackage{appendix}
\newcommand{\qo}[1]{``#1''}                               		
\newcommand{\braket}[2]{\langle#1|#2\rangle}  		
\newcommand{\ket}[1]{|#1\rangle}                      		
\newcommand{\bra}[1]{\langle #1|}                     		
\newcommand{\abs}[1]{\left| #1 \right|}			
\newcommand{\beq}{\begin{equation}}
\newcommand{\eeq}{\end{equation}}
\newcommand{\bei}{\begin{itemize}}			
	\newcommand{\eei}{\end{itemize}}			

\newcommand{\vectgr}[1]{{\boldsymbol#1}}    		
 
\usepackage{hyperref}
\hypersetup{%
   pdfpagemode=None, 
   pdfstartpage=1,
   pdfmenubar=true,
   pdftoolbar=true,
   colorlinks = true,
   linkcolor=blue,
   citecolor=blue,
   urlcolor=blue,
   bookmarksopen=false
 }

\draft 

\begin{document}


\title{Bloch-Landau-Zener dynamics induced by a synthetic field in a photonic quantum walk} 

\author{Alessio D'Errico}\email{aderrico@uottawa.ca}
\affiliation{Dipartimento di Fisica, Universit\`{a} di Napoli Federico II, Complesso Universitario di Monte Sant'Angelo, Via Cintia, 80126 Napoli, Italy}
\affiliation{Department of Physics, University of Ottawa, 25 Templeton Street, K1N 6N5, Ottawa, ON, Canada}
\author{Raouf Barboza}
\affiliation{Dipartimento di Fisica, Universit\`{a} di Napoli Federico II, Complesso Universitario di Monte Sant'Angelo, Via Cintia, 80126 Napoli, Italy}
\author{Rebeca Tudor}
\affiliation{Dipartimento di Fisica, Universit\`{a} di Napoli Federico II, Complesso Universitario di Monte Sant'Angelo, Via Cintia, 80126 Napoli, Italy}
\affiliation{National Institute for Research and Development in Microtechnologies IMT, Bucharest, 077190, Romania}

\author {Alexandre Dauphin}
\affiliation{ICFO -- Institut de Ciencies Fotoniques, The Barcelona Institute of Science and Technology, 08860 Castelldefels (Barcelona), Spain}
\author {Pietro Massignan}
\affiliation{ICFO -- Institut de Ciencies Fotoniques, The Barcelona Institute of Science and Technology, 08860 Castelldefels (Barcelona), Spain}
\affiliation{Departament de F\'isica, Universitat Polit\`ecnica de Catalunya, Campus Nord B4-B5, 08034 Barcelona, Spain}
\author{Lorenzo Marrucci}
\affiliation{Dipartimento di Fisica, Universit\`{a} di Napoli Federico II, Complesso Universitario di Monte Sant'Angelo, Via Cintia, 80126 Napoli, Italy}
\affiliation{CNR-ISASI, Institute of Applied Science and Intelligent Systems, Via Campi Flegrei 34, 80078 Pozzuoli (NA), Italy}
\author{Filippo Cardano}\email{filippo.cardano2@unina.it}
\affiliation{Dipartimento di Fisica, Universit\`{a} di Napoli Federico II, Complesso Universitario di Monte Sant'Angelo, Via Cintia, 80126 Napoli, Italy}




\date{\today}

\begin{abstract}
Quantum walks are processes that model dynamics in coherent systems. Their experimental implementations proved key to unveil novel phenomena in Floquet topological insulators. Here we realize a photonic quantum walk in the presence of a synthetic gauge field, which mimics the action of an electric field on a charged particle.
By tuning the energy gaps between the two quasi-energy bands, we investigate intriguing system dynamics characterized by the interplay between Bloch oscillations and Landau-Zener transitions. When both gaps at quasi-energy values 0 and $\pi$ are vanishingly small, the Floquet dynamics follows a ballistic spreading.
\end{abstract}

\pacs{}

\maketitle 


\section{\label{sec:level1}Introduction}
Quantum walks (QW) are periodically driven processes describing the evolution of quantum particles (walkers) on a lattice or a graph~\cite{Aharonov1993,Kendon2006}. The walker evolution is determined by the unitary translation operators that, at each timestep, couple the particle to its neighbouring sites, in a way that is conditioned by the state of an internal degree of freedom, referred to as \qo{the coin}. An additional unitary operator acts on the internal degrees of freedom and therefore mimicks the \qo{coin tossing} of the classical random walk, and is usually referred to as \textit{coin rotation}~\cite{Aharonov1993}. Besides the original interest in QWs for quantum computation~\cite{Shenvi2003, Childs2004, Childs2009, Childs2010,Lovett2010}, these processes have proved to be powerful tools to investigate topological systems~\cite{Kitagawa2010a, Kitagawa2012, Zeuner2015, Cardano2016, Cardano2017, Barkhofen2017, Ramasesh2017, Zhan2017, Xiao2017, Chen2018, Wang2018, DErrico2020,DErrico2020b}, disordered systems and Anderson localization~\cite{Schreiber2011, Crespi2013a, Edge2015,Harris2017}, multiparticle interactions and correlations \cite{Peruzzo2010, Schreiber2012, Sansoni2012}. QWs have been implemented in many different physical platforms: atoms in optical lattices~\cite{Karski2009,Genske2013}, trapped ions~\cite{Zahringer2010}, Bose-Einstein condensates~\cite{Dadras2018}, superconducting qubits in microwave cavities~\cite{Ramasesh2017} and photonic setups~\cite{Broome2010, Peruzzo2010,Schrieber2010,Sansoni2012,Cardano2015, Zeuner2015, DErrico2020}. They exhibit peculiar properties when an external force is acting on the walker, mimicking for instance the effect of an electric field on a charged particle. These processes, baptized as \qo{Electric Quantum Walks} in Ref.~\cite{Genske2013}, have been studied theoretically in previous works~\cite{Wojcik2004, Banuls2006} and implemented using neutral atoms in optical lattices~\cite{Genske2013}, photons~\cite{Xue2015} and transmon qubits in optical cavities~\cite{Ramasesh2017}. Recently, these concepts have been generalized to 2D QWs \cite{DErrico2020,Chalabi2020}. In this scenario, the walker dynamics can be remarkably different with respect to a standard QW evolution. While in the absence of a force the walker wavefunction spreads ballistically, by applying a constant force it is possible to observe revivals of the initial distribution at specific timesteps~\cite{Genske2013}, induced by Bloch oscillations. Electric QWs are thus an ideal platform to investigate spatial localization induced by "irrational forces"~\cite{Banuls2006, Genske2013, Cedzich2016}, revivals of probability distributions~\cite{Cedzich2013,Xue2014,Cedzich2016,Bru2016,Nitsche2018} and can be used to detect topological invariants~\cite{Atala2013,Abanin2013,Ramasesh2017,Flurin2017, DErrico2020, Upreti2020}.\\
Revivals in electric QWs may appear as a consequence of Bloch oscillations~\cite{Wojcik2004,Flurin2017,Chalabi2020}, and are observed in the case of forces that are much smaller than the relevant energy gap. Revival effects can be indeed destroyed by Landau-Zener transitions~\cite{Shevchenko2009, Flurin2017}. However, in the continuous time regime, it is well known that the interplay between Landau-Zener transitions and Bloch oscillations manifests itself in processes with two characteristic periods \cite{Dreisow2009} that, under specific circumstances, may lead to breathing phenomena even when interband transitions are not negligible. Here we make use of a novel platform which exploits the space of transverse momentum of a paraxial light beam~\cite{DErrico2020} to generate electric QWs, and we observe revivals due to either Bloch oscillations or multiple Landau-Zener transitions. Finally, we discuss how the Floquet nature of these systems affects the walker dynamics. In particular, when the two energy gaps in the spectrum are sufficiently small, the number of LZ transitions is doubled within a single period. This in turn causes the appearance of multiple trajectories, arranged in peculiar regular patterns. 
\begin{figure*}[t!]
\centering
\vskip 0 pt
\includegraphics[scale=0.5]{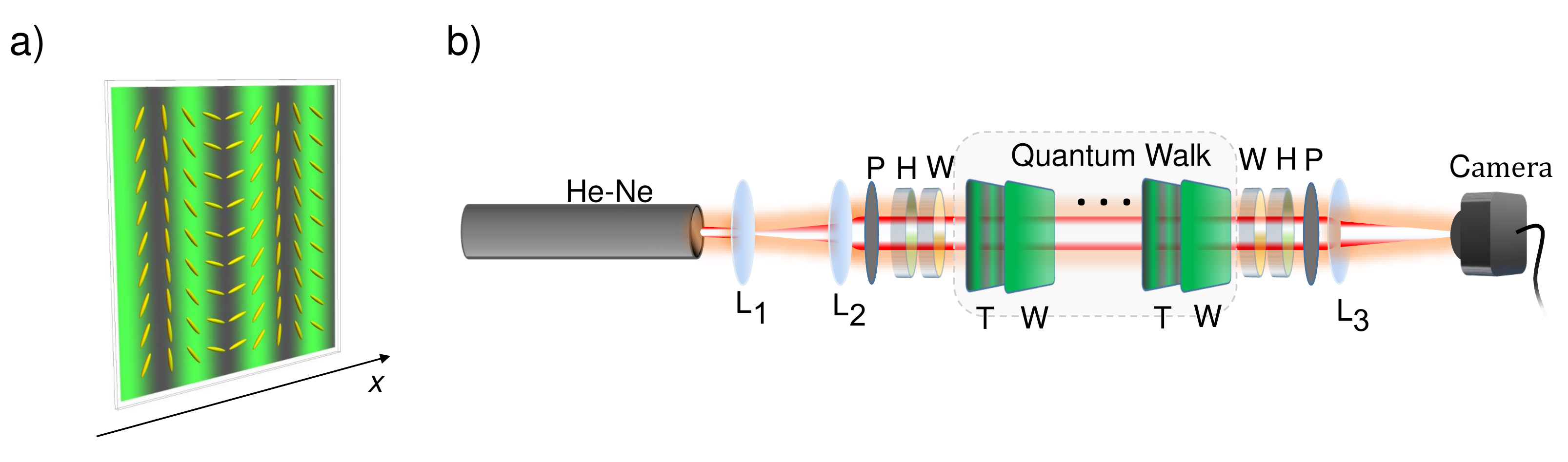}
\caption{{\bf LC cells and experimental apparatus.} a) Schematic illustration of a $g$-plate. The ellipsoids show the local orientation of the liquid crystal molecules; the background pattern corresponds to the intensity of white light transmitted by crossed polarizers, when a $g$-plate is placed between them. b) Sketch of the experimental apparatus. A He-Ne laser beam is expanded by using two lenses (L$_1$ and L$_2$). The input polarization is adjusted by using a polarizer (P), a half-waveplate (H) and a quarter-waveplate (W). The QW is performed by alternating $g$-plates (T) and quarter waveplates (W). An optical 2D Fourier transform is performed on the final state by a converging lens. In its focal plane, the light intensity distribution is recorded by a camera. An additional stage (W-H-P) can be inserted before lens L$_3$ to analyze a single polarization/coin component.}
\label{fig:exp}
\end{figure*}
\section{\label{sec2}Quantum walk in the momentum space of light}
To simulate a QW on a one-dimensional lattice we encode the walker into the transverse momentum of a paraxial light beam. In particular, the walker position, that is identified by an integer coordinate $m$, is associated with the photonic spatial mode $\ket{m}$ given by: 
\begin{align}\label{eq:gaussianbeamtilted}
\ket{m}=A(x,y,z) e^{i[(\Delta k\,m)x+k_z z]},
\end{align} 
where $k_z$ is the wavevector component along the $z$ direction, $\Delta k$ is a constant such that $\Delta k\ll k_z$ and $A$ is a Gaussian spatial envelope with a beam waist $w_0$. Modes described in Eq.\ \eqref{eq:gaussianbeamtilted} are standard Gaussian beams, propagating along a direction that is slightly tilted with respect to the $z$-axis.

At each time step, the evolution is performed by the successive application of a rotation $W$ of the coin degree of freedom, and a translation $T$ that shifts the walker to the left or to the right depending on the coin state being $\ket L$ or $\ket R$. In our setup, both operators are implemented by liquid-crystals (LC) birefringent waveplates. Along these plates the LC molecular orientation angle $\alpha$ is suitably patterned \cite{Rubano2019}, as shown for instance in Fig.\ \ref{fig:exp}a). Here, the operator $T$ is obtained when the local orientation $\alpha$ of the optic axis increases linearly along $x$
\begin{align}\label{eq:Lambda}
\alpha(x,y)=\pi x/\Lambda+\alpha_0,
\end{align}
where $\Lambda$ is the spatial periodicity of the angular pattern and $\alpha_0$ is a constant, thereby forming a regular pattern reminiscent of a diffraction grating. This device has been originally named $g$-plate \cite{DErrico2020}. In the basis of circular polarizations $\ket{L}=(1,0)^T$ and $\ket{R}=(0,1)^T$, the associated operator can be written as
\beq\label{eq:gplates}
T\equiv
\left( {\begin{array}{cc}
   \cos(\delta/2) & i \sin(\delta/2) e^{-2i\alpha_0}\hat t \\
   i \sin(\delta/2) e^{2i\alpha_0}\hat t^\dagger & \cos(\delta/2) \\
  \end{array} } \right),
\eeq
where $\hat t$ and $\hat t^\dagger$ are the (spin-independent) left and right translation operators along $x$, acting respectively as $\hat t\ket{m,\phi}=\ket{m-1,\phi}$ and $\hat t^\dagger \ket{m,\phi}=\ket{m+1,\phi}$ on the spatial modes in Eq.\ \eqref{eq:gaussianbeamtilted}.  Here $\ket{\phi}$ is a generic polarization state, and $\delta$ is the LC optical retardation which can be tuned by adjusting the amplitude of an alternating voltage applied to the cell \cite{Piccirillo2010}. The coin rotation is realized by uniform LC plates ($\alpha=0$), represented by the operator 
\beq\label{eq:L}
L(\delta)=
\left( {\begin{array}{cc}
   \cos(\delta/2) & i\, \sin(\delta/2) \\
   i\,\sin(\delta/2) & \cos(\delta/2) \\
  \end{array} } \right).
\eeq
Typically, we set $\delta=\pi/2$, so as to obtain a standard quarter-wave plate $W=L(\pi/2)$. The quantum walk is realized by applying repeatedly the single step unitary process
\begin{equation}\label{eq:u1}
U_0(\delta, \alpha_0)=T(\delta, \alpha_0)\cdot W,
\end{equation}
and the state after $t$ steps is given by $\ket{\psi(t)}=U_0^t\ket{\psi(0)}$.
Relying on this approach, we realize our QW in the setup sketched in Fig.~\ref{fig:exp}b). A coherent light beam (produced by a He:Ne laser source, with wavelength $\lambda=633$ nm), whose spatial envelope is that of a Gaussian mode, is initially expanded to reach a waist $w_0\simeq 5$ mm. After preparing the desired beam polarization with a polarizer~($P$), a half-waveplate~($H$) and a quarter-waveplate~($W$), we perform the quantum walk by letting the beam pass through a sequence of $g$-plates ($T$) and quarter-waveplates. In the present experiment we realized walks containing 14 unit steps. 
The optical retardation of the $g$-plates is controlled by tuning an alternating voltage, and their spatial period is $\Lambda=w_0=5$ mm. As explained in detail in Ref.~\cite{DErrico2020}, with this choice of parameters we simulate the evolution of an initial state that is localized in the transverse wavevector space, corresponding to the spatial mode $\ket m=0$. All the devices that implement the QW are liquid crystal plates, fabricated in our laboratories and mounted in a compact setup. The final probability distribution is extracted from the intensity distribution in the focal plane of a converging lens located at the end of the QW~\cite{DErrico2020}. This distribution consists of an array of Gaussian spots, centered on the lattice sites, whose relative power (normalized with respect to the total power) give the corresponding walker probabilities. An additional set of waveplates and a polarizer can be placed before the lens to analyze specific polarization components. We use these projections to prove that, when a substantial revival of the probability distribution is observed, the coin part of the final state corresponds to the initial one.

\section{\label{sec_EQW} Realizing an electric QW}

The discrete translational symmetry in the walker space implies that the unitary single step operator $U_0$ can be block-diagonalized in the quasi-momentum basis~\cite{Kitagawa2010a}
\begin{equation}
U_0=\int_{-\pi}^{\pi}\frac{dq}{2\pi}\,\mathcal{U}_0(q)\otimes \ket{q}\bra{q},
\label{eq:diagU}
\end{equation}
where $\ket{q}=\sum_m e^{iqm}\ket{m}$, with $q$ varying in the first Brillouin zone $BZ=[-\pi,\pi)$. In the case of a 2D coin space, the operator $\mathcal{U}_0(q)$ is a $2\times2$ unitary matrix which may be written as
\begin{equation}
\mathcal{U}_0(q)=\text{exp}[-i E(q)\mathbf{n}(q)\cdot\vectgr{\sigma}],
\label{eq:eigenenergies}
\end{equation}
where $\mathbf{n}(q)$ is a unit vector, $\vectgr{\sigma}=(\sigma_1,\sigma_2,\sigma_3)$ is the vector composed of the three Pauli matrices, and $\pm E(q)$ are the quasienergies of the two bands of the system \cite{Kitagawa2010a} [see Fig.\ \ref{fig:simulations}a)]. Being the QW a Floquet evolution, this spectrum exhibits two gaps at quasi-energies values $0$ and $\pi$. For practical reasons, we define the quantity $E_g$ as the minimum value between the two gap sizes (when varying the quasi-momentum in the BZ). In the following we will denote the eigenstates of the quantum walk evolution in the absence of external force as $\ket{u_{\pm}(q)}\otimes\ket{q}$, where  $\ket{u_{\pm}(q)}$ is the coin part.
In this work, we implement QWs corresponding to two different regimes: $\delta=\pi$ and $\delta=\pi/2$. In the first case $E_g$ is at its maximum value, providing the optimal configuration for the observation of clean Bloch oscillations. In the second case the gap at $E=0$ vanishes [see Fig.\ \ref{fig:simulations}a)]. In this case, a LZ transition occurs with unit probability.

As discussed in previous works \cite{Genske2013, DErrico2020}, applying an external constant force $F$ is equivalent to shifting linearly in time the quasi-momentum: $q(t)= q(0)+F t$. Hence the single step operator at the time-step $t$, labelled as $\mathcal{U}(q,t)$, satisfies the equation
\begin{equation}
\mathcal{U}(q,t)=\mathcal{U}_0(q+Ft).
\label{eq:electricdiagU}
\end{equation}
In our setup the walker position is encoded into the optical transverse wavevector. As such, the walker quasi-momentum corresponds to the spatial coordinate $x$ in our laboratory reference frame, introduced to define spatial modes in Eq.\ \eqref{eq:gaussianbeamtilted}. In particular, $x$ and $q$ are related by the following expression \cite{DErrico2020}
\begin{align}
q=-\frac{2\pi x}{\Lambda}.
\label{eq:quasimomentum_position}
\end{align}
Using Eqs.\ \eqref{eq:electricdiagU} and \eqref{eq:quasimomentum_position}, it is straightforward to see that the effect of a constant force is simulated if the $g$-plates corresponding to the timestep $t$ are shifted along $x$ by the amount $\Delta x_t=t\,\Lambda F/(2\pi)$.
\\
\\

\section{\label{sec3}Refocusing effects in electric quantum walks}

\begin{figure}
\includegraphics[width=\columnwidth]{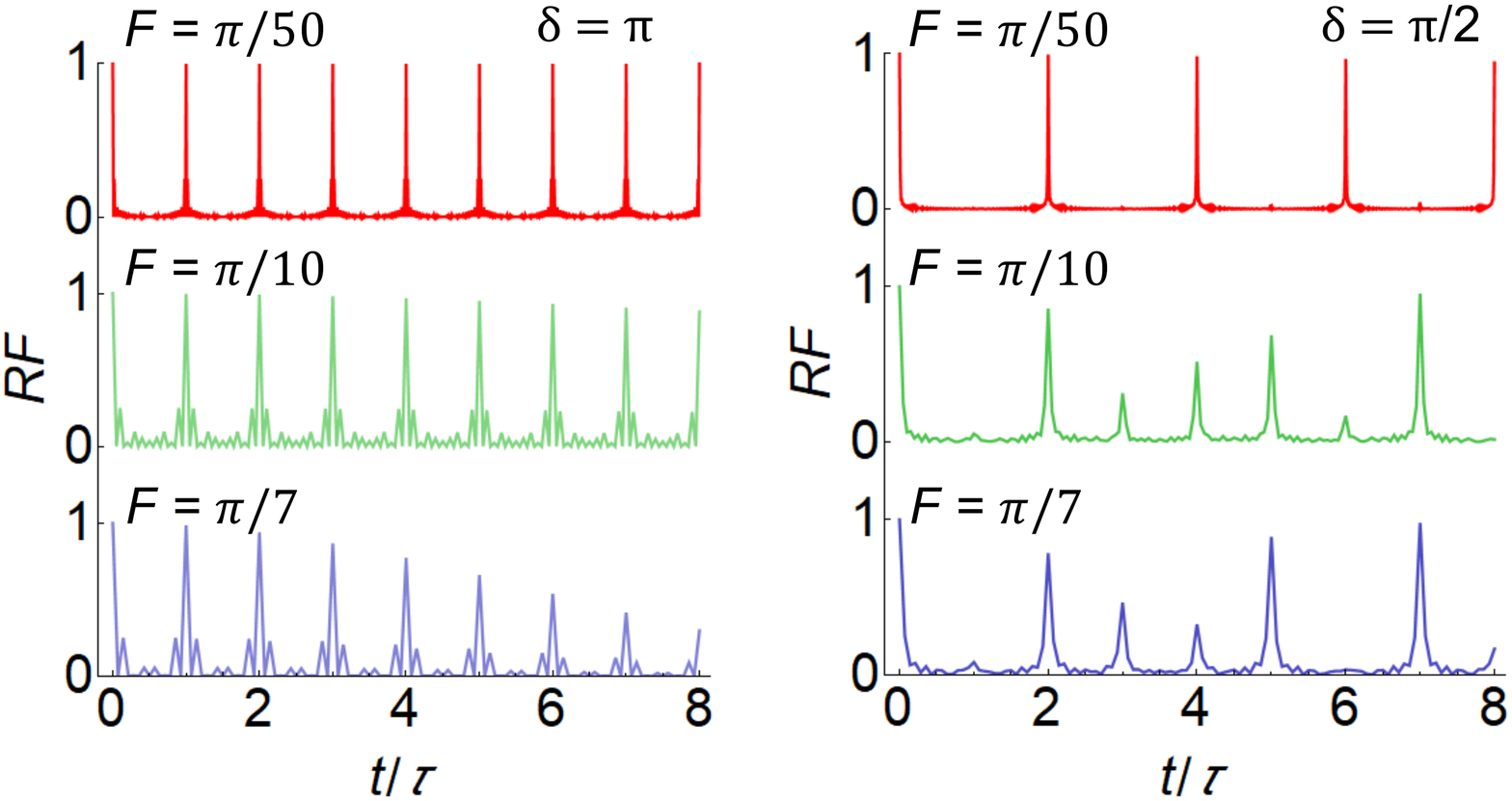}
\caption{{\bf Refocusing fidelity.} Values of the RF calculated during the evolution, for various strengths of the force $F$. In the left column the optical retardation is set to $\delta=\pi$. The large spectral gap inhibits LZ transitions, so that the system performs clean Bloch oscillations with very large RF. In the right column $\delta$ is set to $\pi/2$ so that the spectral gap is closed. Very weak forces (red line) induce an almost perfect refocusing at twice the Bloch period $\tau$. 
}
\label{fig:RF}
\end{figure}

Bloch oscillations have been extensively studied in the continuous time regime (see, e.g., Refs.~\cite{Hartmann2004, Dominguez-Adame2010}) and have been recently considered in discrete time settings~\cite{Cedzich2013, arnault2020}. Here we review the theory of Bloch oscillations and Landau-Zener transitions, and investigate their phenomenology in our quantum walk protocol. \\
\begin{figure*}[!t]
\centering
\vskip 0 pt
\includegraphics[width=\textwidth]{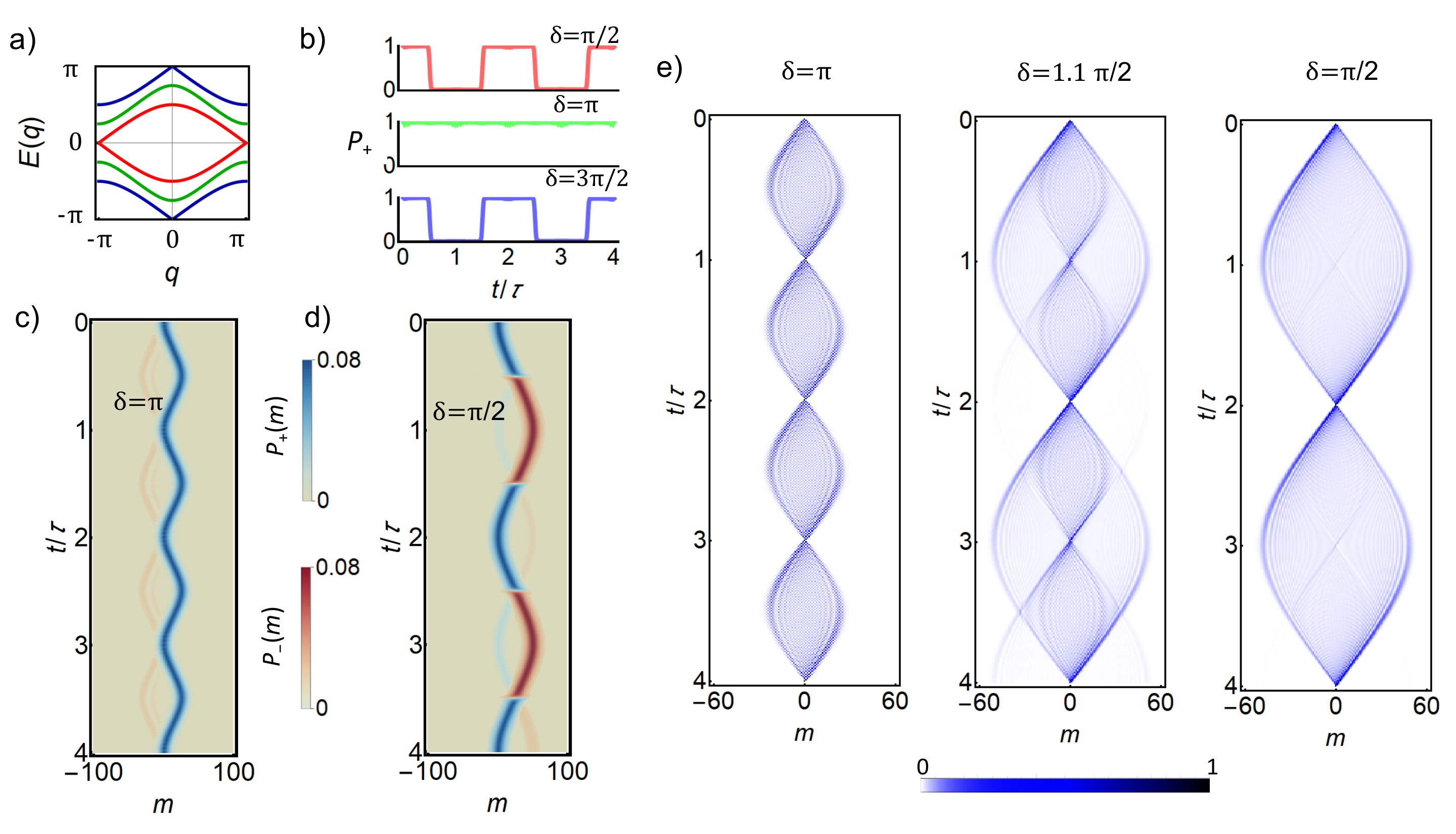}
\caption{{\bf Bloch oscillations in electric quantum walks.} a) Quasi-energy dispersion $E(q)$ for various optical retardations: $\delta=\pi/2$ (red), $\delta=\pi$ (green) and $\delta=3\pi/2$ (blue). b-d) We simulate the evolution of broad wavepackets prepared in the upper band with a width $w=10$, under an external force $F=\pi/50$. In panel b), we plot the probability that the evolved state is located on the upper band. Panels c) and d) show the the spatial probability distributions relative to the two bands ($P_+(m)$ and $P_-(m)$) as a function of the timestep. 
e) Evolutions of initially localized states generated during the QW dynamics under a force $F=\pi/50$, for three different values of delta. Localized inputs give rise to trajectories with multiple revivals. The initial state of these simulations is $\ket{H}\ket{m=0}=(\ket{R}+\ket{L})\ket{0}/\sqrt{2}$. 
}
\label{fig:simulations}
\end{figure*}
Consider the evolution of an initial state that approximates an eigenstate of the system without any force, namely a Gaussian wavepacket centered around $m=0$
\begin{equation}\label{wavepacket}
\ket{\Psi(q_0,w)}=\mathcal{N}\sum_m \exp[-(m/w)^2+i q_0 m]\ket{u_{\pm}(q_0)}\otimes\ket{m},
\end{equation}
where $\mathcal{N}$ is a normalization factor and $w$ controls the width of the wave packet. We consider wavepackets with large values of $w$, so that these states are sharply peaked in quasi-momentum space. If the external force $F$ is small with respect to the minimum energy gap, the adiabatic approximation dictates that (i) the wavepacket remains in the original quasi-energy band during the whole evolution, (ii) the coin state rotates as $\ket{u_{\pm}(q_0)}\rightarrow\ket{u_{\pm}(q_0+Ft)}$ and (iii) the center of mass follows the equation of motion: $m(t)=(1/F)\int_{q_0}^{q_0+Ft}v_g(q) dq$, where $v_g(q)=\partial_q E_\pm(q)$ is the group velocity. In particular, after a period $\tau=2\pi/F$, the wavepacket gets back to the original state. This result is illustrated in Figs. \ref{fig:simulations} b,c) for the case in which the initial state is in the upper band. In panel b) we plot the probability that the walker is found in the upper energy band, which remains approximately equal to one across the evolution. In panel c), we report in a single plot the probability $P_\pm(m,t)$ that, at the time-step $t$, the walker is found on the lattice site $m$ in the upper (lower) band.  A different scenario occurs for a closed energy gap (as for instance at $\delta=\pi/2$). In this case the adiabatic approximation breaks up when the wavepacket reaches the region of the Brillouin zone where $E_g$ is minimum. The Landau-Zener theory \cite{Shevchenko2009} predicts that the transition to the lowest band occurs with unit probability in the case of zero gap. We clearly observe this phenomenon in our simulations [see Fig. \ref{fig:simulations}b,d)]. After a period $\tau$, the the wavepacket is entirely found in the lowest band, as a consequence of a LZ transition. However, at $t=2\tau$, a second transition takes place and the input state is restored. Remarkably, the same dynamics is observed for $\delta=3\pi/2$ where the gap between the two bands vanishes at $E(q=0)=\pm \pi$. Such situation can only appear in a Floquet system.\\
Observing the dynamics of Gaussian wavepackets facilitates the understanding of the evolution of localized input states, that are not confined to a single band. In this situation, the oscillating behavior of the system eigenstates can lead to a refocusing of the localized input. To illustrate this result, let us consider a generic initial state:
\begin{equation}
\ket{\psi_0}=\int_{-\pi}^{\pi}\frac{dq}{2\pi}\left[c_+(q)\ket{u_+(q)}+c_-(q)\ket{u_-(q)} \right]\otimes\ket{q}.
\end{equation} 
The state after $\tau$ steps is given by 
\begin{align}\label{psitau}
\ket{\psi_\tau}=&\int_{-\pi}^{\pi}\frac{dq}{2\pi}\prod_{t=0}^{\tau}\mathcal{U}_0(q+Ft)\nonumber \\
                &\times\left[c_+(q)\ket{u_+(q)}+c_-(q)\ket{u_-(q)} \right]\otimes\ket{q}
\end{align}
where the product $\prod_{t=0}^{\tau}\mathcal{U}_0(q+Ft)$ must be written in the time ordered form. In the adiabatic regime where $F\ll E_g$ we find (see Appendix \ref{AppA} for details)
\begin{align}\label{Eq.Nstep}
\ket{\psi_\tau}=\int_{-\pi}^{\pi}\frac{dq}{2\pi}\bigr(&c_+(q)e^{-i(\gamma_+ +\gamma_g)}\ket{u_+(q)}\nonumber\\
&+c_-(q)e^{-i(\gamma_-  +\gamma_g)}\ket{u_-(q)} \bigr)\otimes\ket{q}.
\end{align}
The Zak phase $\gamma_g$ \cite{Zak1989} appears as a global phase, and does not play an important role here. We rather focus our attention on the dynamical phases acquired by the eigenstates when undergoing a complete Bloch oscillation, that are given by
\begin{equation}
\gamma_{\pm}=\sum_{j=0}^{(2\pi/F-1)}E_{\pm}(q+jF).
\end{equation}
These phases are independent of $q$ in the limit of small $F$, hence they can be factored out from the integrals. Therefore, if at $t=0$ only one band is occupied, the final state concides with the initial one, apart from a global phase factor. When the system is initially prepared in a state occupying both bands, complete refocusing can be observed when the difference between the dynamical phases, $\Delta\gamma=\gamma_+-\gamma_-$, is a multiple of $2\pi$. In general, the final state will be different from the initial one, due to the additional relative phase acquired by the states over the two bands. In particular, for $\Delta\gamma=\pi$, the final state is orthogonal to the initial one, even though it is still localized at the initial lattice site.\\
At $\delta=\pi$, the difference $\Delta\gamma$ is $2\pi$ for $ F=2\pi/2l$, with $l$ integer, and $\pi$ for $ F=2\pi/(2l+1)$. In the first case we can observe refocusing of the full quantum state after a number of steps that is a multiple of $\tau=2\pi/F$, as shown in Fig. \ref{fig:simulations}e). 
For a vanishing gap, this description breaks down in proximity of the gap-closing point. However, also in this case, a refocusing of the input state can be observed at time-steps multiples of $2\tau$, as a result of the even number of Landau-Zener transitions occurring for each eigenstate \cite{Breid2006,Shevchenko2009} [see Fig.\ \ref{fig:simulations}d)]. This is confirmed by the results of numerical simulations reported in Fig.\ \ref{fig:simulations}e) for a QW with $\delta=\pi/2$. The evolution at $\delta=3\pi/2$ is actually identical to the latter case. Indeed, the two single step operators are the same, apart from a global phase factor $\exp(i\pi)$ and a $\pi$ shift of the whole BZ.\\ A quantitative analysis of non-adiabatic effects in QWs with $\delta=\pi$ and $\delta=\pi/2$ is provided in Fig.~\ref{fig:RF}, depicting the refocusing fidelity $\text{RF}=\abs{\braket{\psi(n \tau)}{\psi(0)}}^2$. When $\delta=\pi$, the RF is peaked at $t=n\tau$, and it is approximatively equal to one for small values of the force, while the peak value decreases with increasing $F$. In the case $\delta=\pi/2$, for small values of the force ($F=\pi/50$) a good refocusing is observed at $t=2 n\tau$ (with $n$ integer). For stronger forces (e.g. $F=\pi/10$ and $F=\pi/7$) the behavior at longer times appears more complicated due to the effect of residual Bloch oscillations, responsible for the peaks at odd multiples of $\tau$.
\\
\section{Double LZ transitions within a single oscillation period}
\begin{figure}[!t]
\centering
\vskip 0 pt
\includegraphics[width=\columnwidth]{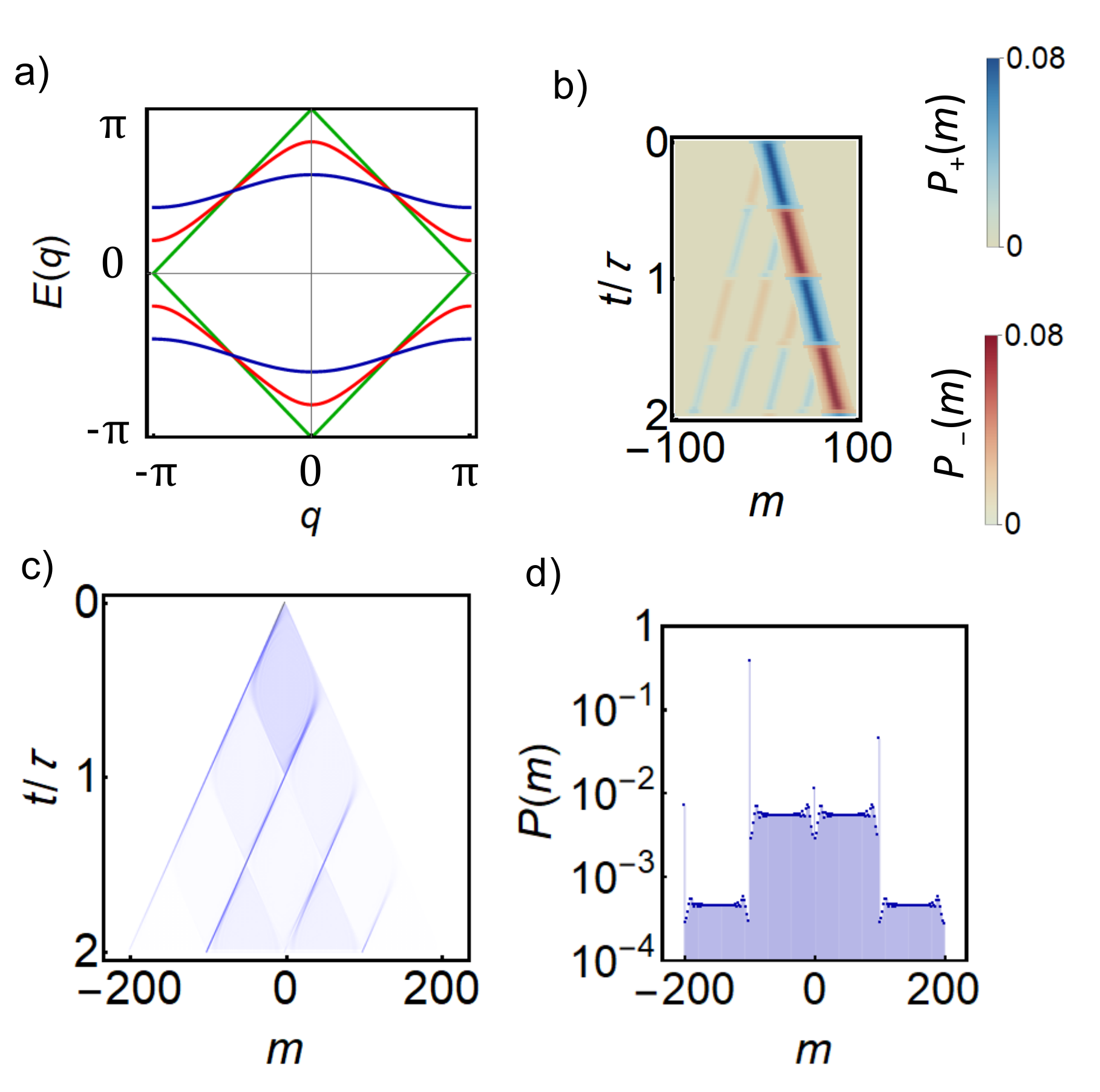}
\caption{\textbf{Double LZ transitions within a single oscillating period.} a) Quasi-energy spectrum of the protocol $U_0(\delta)$ for three values of $\delta$,  $\delta=0.5\pi$ (blue), $\delta=0.7 \pi$ (red), $\delta=0.9 \pi$ (green); b) Gaussian wavepacket evolution [$F=\pi/20$, $\delta=0.9 \pi$]. As in  Fig.\ \ref{fig:simulations} the wavepacket is prepared with a width $w=10$ and coin corresponding to the upper band eigenstate at $q=0$. The plot shows simultaneously the probability distributions relative to the upper and lower band. c) Localized input state [$F=\pi/50$, $\delta=0.9 \pi$, initial coin state: $\ket{H}$]. d) Total probability distribution at $t=200$.}
\label{fig:DLZ}
\end{figure}
Before discussing the experimental results, we illustrate an additional QW dynamics in the very peculiar case where both energy gaps at $E=0$ and $E=\pi$ are made small, so that LZ can take place with high probability twice when crossing the Brillouin zone. This is a unique feature of Floquet systems. We consider a protocol defined by the single step operator $U_0(\delta)=T(\pi)L(\delta)$. In Fig.\ \ref{fig:DLZ}a) we plot the quasi-energy spectrum for three values of $\delta$. The minimum energy gaps around $E=0$ and $E=\pi$ are located at quasi-momentum values $q=0$ and $q=\pi$, respectively, and have the same amplitude. Their value can be tuned by adjusting $\delta$. Fig.\ \ref{fig:DLZ}b) depicts the walker evolution in the case $\delta=0.9 \,\pi$, considering as input state a wavepacket entirely localized on the upper band. It is clear that within a single period $\tau$ two LZ transitions take place. The fraction of the wavepacket that undergoes two transitions keeps moving in the same direction, as its group velocity does not change sign. Fig.\ \ref{fig:DLZ}c) shows the evolution in the case of a localized input state. Also in this case, at each transition the wavepacket splits in a component that keeps moving in the same direction, and another that is reflected, similarly to a beam splitter. The interplay between these two mechanisms gives rise to a complex dynamics, where at each period $\tau$ the wavepacket is concentrated on a set of lattice sites that are equally spaced, as shown in Fig.\ \ref{fig:DLZ}d).
\section{Experimental results}\label{sec4}
\begin{figure}[t!]
\centering
\vskip 0 pt
\includegraphics[width=\columnwidth]{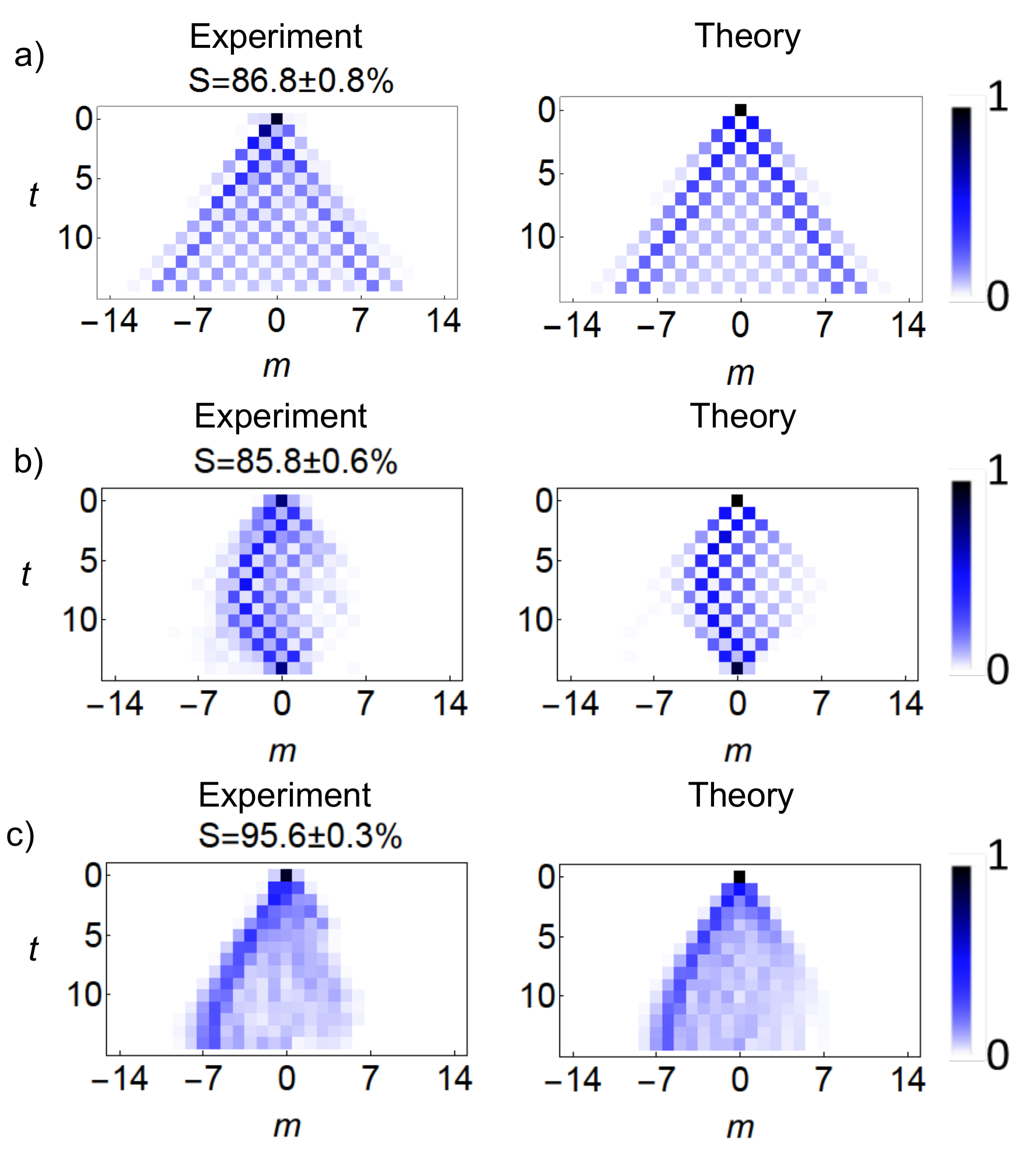}
\caption{{\bf Experimental demonstration of revivals in Electric QWs.} We plot the walker probability distributions of electric quantum walks (14 steps) with or without an applied force. Experimental and theoretical results are reported in the left and right column, respectively. A quantitative comparison is provided in terms of the similarities. a) $\delta=\pi$, $F=0$. b) $\delta=\pi$ (energy gap is maximum), $F=\pi/7$. c) $\delta=\pi/2$ (energy gap is vanishing), $F=\pi/7$. The coin input state is $\ket{\psi(0)}=\ket{H,m=0}=(\ket{L}+\ket{R})\ket{0}/\sqrt{2}$. Experimental data are averages over 4 repeated measurements.}
\label{fig:deltap}
\end{figure}
\begin{figure}[h!]
\centering
\vskip 0 pt
\includegraphics[scale=0.25]{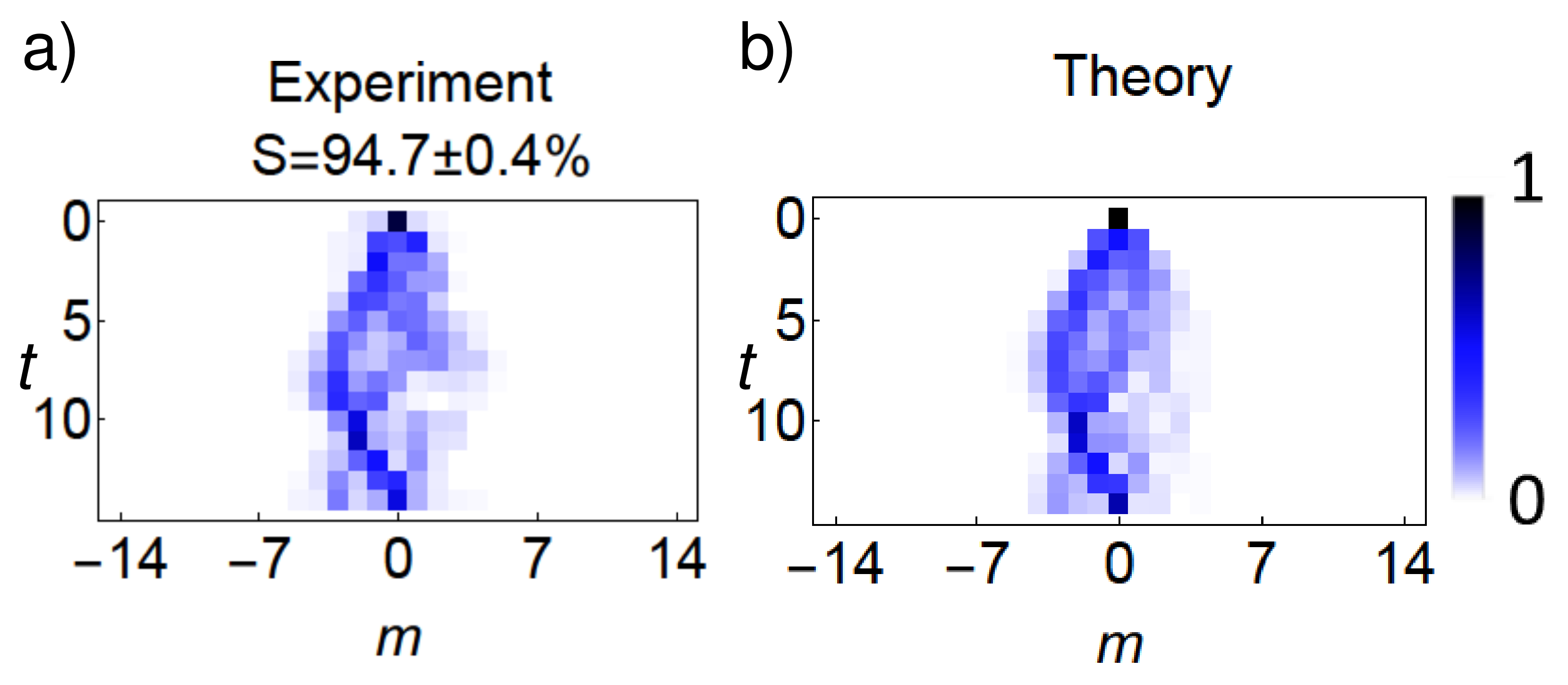}
\caption{{\bf Nonadiabatic refocusing.} We realize 14 steps of an electric quantum walk with force $F=2\pi/7$, such that the Brillouin Zone is explored twice. At $\delta=\pi/2$, due to the gap closing at $q=\pi$, Landau-Zener transitions occur with approximately unit probability. Hence, a double LZ transition should bring the particle to its original state. This is confirmed by our experiments (panel a)) which are in good agreement with theoretical simulations (panel b)). Plots are shown for an $H$-polarized input state. Experimental data are averages over 4 repeated measurements.}
\label{fig:2p7_deltap2}
\end{figure}

We benchmark our platform by first performing experiments of QWs without an external force. The results are reported in Fig.~\ref{fig:deltap}a), showing the well known ballistic propagation that characterizes these processes \cite{Kempe2003}. Experimental probability distributions $\vert\psi_E(m)\vert^2 $ are compared with theoretical simulations $\vert\psi_T(m)\vert^2$, with their agreement being quantified by the similarity \cite{Kailath1967} 
\begin{equation}
S=\sqrt{\sum_m \vert\psi_E(m)\vert^2 \vert\psi_T(m)\vert^2},
\end{equation}
where we are assuming that $\vert\psi_E(m)\vert^2$ and $\vert\psi_T(m)\vert^2$ are normalized. In Figs.\ \ref{fig:deltap},\ref{fig:2p7_deltap2} we report the similarity $S$ averaged over the results for each step. All the experimental errors were obtained by repeating each experiment 4 times.

To confirm experimentally the results discussed in the previous sections, in Fig.\ \ref{fig:deltap}b) we show an electric quantum walk characterized by a force $F=\pi/7$. Here, a complete refocusing is observable at the time-step $t=14$, corresponding to the last step of our evolution. The evolution corresponds to the case $\delta=\pi$, so that the force is still smaller than the energy gap, even if interband transitions are not completely negligible.   Besides being mostly localized at $m=0$, the final state is expected to have the same polarization of the input beam, since refocusing happens after an even number of steps. Defining $\ket{\phi_{\tau}}$ as the polarization state measured for the optical mode with $m=0$ after $\tau$ steps, we calculated the coin refocusing fidelity $R=\abs{\braket{\phi_0}{\phi_{\tau}}}^2$, (not to be confused with the refocusing fidelity of the whole quantum state $RF$ considered in Fig.\ \ref{fig:RF}), that measures the overlap between the two coin states. We obtain $R>98\%$ for three different input states, as shown in Appendix \ref{AppB}. This is in agreement with the adiabatic model we developed earlier, where we showed that, for $F=2\pi/\tau$ with $\tau$ being an even number, the initial state is fully reconstructed after $\tau$ steps.

The contribution of Bloch oscillations is completely suppressed for $\delta=\pi/2$, where the energy spectrum presents a gap closing point and the revival of the input state cannot occur. This is shown in the evolution depicted in Figs.~\ref{fig:deltap}c). After 28 steps an approximate refocusing is expected due to double Landau-Zener transitions (see Sec.~\ref{sec3}). To observe this effect in our apparatus, we doubled the value of the force, that is $F=2\pi/7$. The results are shown in Fig.~\ref{fig:2p7_deltap2}. After 14 steps, the wavefunction is again sharply peaked at the origin, with some broadening in agreement with numerical simulations. Refocusing in the regime of LZ transitions has been observed in the continuous time domain in bent waveguide arrays~\cite{Dreisow2009} but not in a Floquet systems like ours. 
\section{Conclusions}
In this work we studied electric QWs in one spatial dimension, relying on a platform where the walker degree of freedom is encoded in the transverse wave vector of a paraxial light beam~\cite{DErrico2020}. The presence of an external force is mimicked by applying a step-dependent lateral displacement of the liquid-crystal plates, increasing linearly with the step number. By tuning the energy gap of our system we studied experimentally the interplay between Bloch oscillations and Landau-Zener transitions, which influence the emerging of revival of the input distribution of the walker wave packet. This allowed us to show experimentally that Landau-Zener oscillations, i.e. a revival of the probability distribution due to multiple Landau Zener tunneling (already demonstrated in the continuous-time~\cite{Dreisow2009}) can be observed in discrete-time processes. Moreover, we investigated a regime with two transitions in a single Bloch oscillation, by tuning the two gaps of our QW at $E=0, \pi$, which is a possibility unique of Floquet systems. We plan to extend these studies to the regime of irrational forces and to study their interplay with static and dynamic disorder. Moreover, we aim to realize similar experiments in a two dimensional system with the same technology~\cite{DErrico2020}.

\section*{Data Availability}
The data that support the findings of this study are available from the corresponding authors upon reasonable request.

\begin{acknowledgments}
The authors wish to thank M.\ Maffei e C.\ Esposito for their valuable help in the early stage of this project. A.D'E., R.B., L.M and F.C. acknowledge financial support from the European Union Horizon 2020 program, under European Research Council (ERC) grant no. 694683 (PHOSPhOR). A.D'E. acknowledges support by Ontario’s Early Research Award (ERA), Canada Research Chairs (CRC), and Canada First Research Excellence Fund (CFREF). R.T acknowledges financial support from IMT-Bucharest Core Program “MICRO-NANO-SIS-PLUS”  14N/2019, project PN 19160102 funded by MCE and from a grant of the Romanian Ministry of Research and Innovation, PCCDI- UEFISCDI, project number PN-III-P1-1.2-PCCDI-2017- 0338/79PCCDI/2018, within PNCDI III. A.D. acknowledges support from ERC AdG NOQIA, Spanish Ministry of Economy and Competitiveness (Severo Ochoa program for Centres of Excellence in RD (CEX2019-000910-S), Plan National FISICATEAMO and FIDEUA PID2019-106901GB-I00/10.13039 / 501100011033, FPI), Fundacio Privada Cellex, Fundacio Mir-Puig, and from Generalitat de Catalunya (AGAUR Grant No. 2017 SGR 1341, CERCA program, QuantumCAT U16-011424, co-funded by ERDF Operational Program of Catalonia 2014-2020), MINECO-EU QUANTERA MAQS (funded by State Research Agency (AEI) PCI2019-111828-2 / 10.13039/501100011033), EU Horizon 2020 FET-OPEN OPTOLogic (Grant No 899794), and the National Science Centre, Poland-Symfonia Grant No. 2016/20/W/ST4/00314.  A.D. acknowledges the financial support from a fellowship granted by la Caixa Foundation (ID 100010434, fellowship code LCF/BQ/PR20/11770012). P.M. acknowledges financial support from the Spanish MINECO (FIS2017-84114-C2-1-P) and the Generalitat de Catalunya (project QuantumCat, Ref.~001-P-001644).
\end{acknowledgments}

\appendix
\section{Derivation of Eq. (\ref{Eq.Nstep})}\label{AppA}

The operators $\mathcal{U}(q+Ft)$ can be decomposed as
\begin{equation}
\mathcal{U}_0(q_t)=e^{iE(q_t)}\ket{u_+(q_t)}\bra{u_+(q_t)}+e^{-iE(q_t)}\ket{u_-(q_t)}\bra{u_-(q_t)},
\end{equation}
where we have defined $q_t=q+Ft$. In the adiabatic approximation we have $\braket{u_{\pm}(q_{t+1})}{u_{\mp}(q_{t})} \ll 1$, i.e. interband transitions between successive steps happen with low probability. Within this approximation, the whole unitary evolution can be approximated by retaining the terms up to the first order in $F$.
\begin{widetext}
\begin{align}\label{Eq.forceaction}
\prod_{t=0}^{\tau}\mathcal{U}_0(q+Ft)= &\,\left(e^{iE(q_\tau)}\ket{u_+(q_\tau)}\bra{u_+(q_\tau)}+e^{-iE(q_\tau)}\ket{u_-(q_\tau)}\bra{u_-(q_\tau)}\right)\times\ldots\nonumber\\&\times \left(e^{iE(q_0)}\ket{u_+(q_0)}\bra{u_+(q_0)}+e^{-iE(q_0)}\ket{u_-(q_0)}\bra{u_-(q_0)}\right)\nonumber\\
                              \approx&\, e^{i\sum_{t=0}^\tau E(q_t)}\ket{u_+(q_\tau)}\braket{u_+(q_{\tau})}{u_+(q_{\tau-1})}\times\ldots\times\braket{u_+(q_{1})}{u_+(q_{0})}\bra{u_+(q_0)}\nonumber\\
															&+e^{-i\sum_{t=0}^\tau E(q_t)}\ket{u_{-}(q_\tau)}\braket{u_-(q_{\tau})}{u_-(q_{\tau-1})}\times\ldots\times\braket{u_-(q_{1})}{u_-(q_{0})}\bra{u_-(q_0)}+ U_{LZ}\nonumber\\
															=&\,e^{i\sum_{t=0}^\tau E(q_t)}\ket{u_+(q_0)}\bra{u_+(q_0)}\left(\prod_{t=1}^{\tau}\braket{u_+(q_{t})}{u_+(q_{t-1})}\right)\nonumber\\
															&+e^{-i\sum_{t=0}^\tau E(q_t)}\ket{u_-(q_0)}\bra{u_-(q_0)}\left(\prod_{t=1}^{\tau}\braket{u_-(q_{t})}{u_-(q_{t-1})}\right)+ U_{LZ},
\end{align}
\end{widetext}
where we used $q_0=q_{\tau}$. In the limit of small $F$, the contributions $\left(\prod_{t=1}^{\tau}\braket{u_{\pm}(q_{t})}{u_{\pm}(q_{t-1})}\right)$ are equal to $\exp(i\gamma_z^{\pm})$, where $\gamma_z^{\pm}$ are the Zak phases \cite{Berry1984,Zak1989} associated to the single energy bands. In our QW $\gamma_z^{+}=\gamma_z^{-}=\gamma_z$. The term $U_{LZ}$ is an $O(F)$ contribution related to interband transitions. In the following, we show that its amplitude is negligible for $F\ll E_g$, where $E_g$ is the energy gap defined in the main text (see also Ref. \cite{Ramasesh2017}).
Let us evaluate explicitly $U_{LZ}$ in order to show that it is negligible in the adiabatic approximation. $U_{LZ}$ is given, to order $O(F)$, by a sum, $U_{LZ}=\sum_{t^*=0}^{\tau}U_{LZ}(t^*)$, where $U_{LZ}(t^*)$ describes a process where a single Landau-Zener transition happens at $q=q_{t^*}$:
\begin{align}
          U_{LZ}(t^*)=&\ket{u_+(q_0)}\bra{u_-(q_0)}\nonumber\\
					            &\times\left(\prod_{t=t^*+1}^{\tau}\braket{u_{+}(q_{t+1})}{u_{+}(q_t)}e^{-iE(q_t)}\right) \nonumber \\
                    &\times \braket{\partial_q u_+(q_{t^*})}{u_-(q_{t^*})}F\nonumber\\
										&\times\left(\prod_{t=0}^{t^*-1}\braket{u_-(q_{t+1})}{u_{-}(q_t)}e^{iE(q_t)}\right).
\end{align}
The $O(F)$ contributions to $U_{LZ}$ are obtained setting $\braket{u_{\pm}(q_{t+1})}{u_{\pm}(q_t)}\approx 1$:
\begin{align}
U_{LZ}=&\ket{u_+(q_0)}\bra{u_-(q_0)}\nonumber\\
					            &\times\left(\sum_{t^*=0}^{\tau}e^{-i\gamma_+}e^{2i\sum_{t=0}^{t^*} E(q_t)}\braket{\partial_q u_{+}(q_{t^*})}{u_{-}(q_{t^*})}\right)F\nonumber\\
					            &+(u_+\leftrightarrow u_-).
\end{align}
From this result it follows that the probability of a Landau-Zener transition is
\begin{equation}\label{eq:PLZ}
P_{+-}=F^2\abs{\sum_{t^*=0}^{\tau}e^{2i\sum_{t=0}^{t^*} E(q_t)}\braket{\partial_q u_{+}(q_{t^*})}{u_{-}(q_{t^*})}}^2.
\end{equation}
This formula corresponds to the one found in Ref. \cite{Ramasesh2017}. Within the approximations considered ($F\ll E_g$), $P_{+-}$ is negligible (see also Ref. \cite{Ramasesh2017} for split-step quantum walk protocols). In particular we have $P_{+-}<0.1$ at $\delta=\pi$ for $F<\pi/7$. In this regime we can thus discard the term $U_{LZ}$ and substituting Eq.\ \eqref{Eq.forceaction} in Eq.\ \eqref{psitau} we obtain Eq. \eqref{Eq.Nstep}.

\section{Experimental data for the evaluation of refocusing fidelity}\label{AppB}
To show that, for $\delta=\pi$, the measured central spot of the final probability distribution has the same polarization as the input state, we measured the projections on the three mutually unbiased bases: $\{\ket{H}, \ket{V}\}$, $\{\ket{A}, \ket{D}\}$ and $\{\ket{L}, \ket{R}\}$, which allows to reconstruct the full coin state. Here $\ket{A,D}=(\ket{L}\mp i \ket{R})/\sqrt{2}$, and $\ket{V}=i(\ket{L}-\ket{R})/\sqrt{2}$. We obtained $R=97.8\pm2\%$ for $\ket{\psi(0)}=\ket{H,m=0}$, $R=98.5\pm2\%$ for $\ket{\psi(0)}=(\ket{L}-i \ket{R})\ket{0}/\sqrt{2}$, and $R=99\pm3\%$ for $\ket{\psi(0)}=\ket{L}\ket{0}$. Fidelities are calculated by using intensities recorded in a square of $4\times 4$ pixels centered on the maximum of each spot.
\providecommand{\noopsort}[1]{}\providecommand{\singleletter}[1]{#1}%

\end{document}